\documentclass[twocolumn,showpacs,preprintnumbers,amsmath,amssymb]{revtex4}

\usepackage{graphicx}
\usepackage{dcolumn}
\usepackage{bm}
\newcommand{\Xstate}{X~$^1\Sigma_{\mathrm{g}}^+$}
\newcommand{\astate}{a~$^3\Sigma_{\mathrm{u}}^+$}
\newcommand{\wn}{cm$^{-1}$}

\begin{document}

\preprint{}

\title{Potassium ground state scattering parameters and \\
Born-Oppenheimer potentials from molecular spectroscopy}%

\author{Stephan Falke}
\author{Horst Kn\"ockel}
\author{Jan Friebe}
\author{Matthias Riedmann}
\author{Eberhard Tiemann}
\affiliation{Institut f\"ur Quantenoptik, Leibniz Universit\"at Hannover, Welfengarten~1, 30167 Hannover, Germany}%

\author{Christian Lisdat}
 \email{christian.lisdat@ptb.de}
\affiliation{Physikalisch-Technische Bundesanstalt, Bundesallee 100, 38116 Braunschweig, Germany}%

\date{\today}

\begin{abstract}
We present precision measurements with MHz uncertainty of the energy gap between asymptotic and well bound levels in the electronic ground state \Xstate\ of the $^{39}$K$_2$ molecule. The molecules are prepared in a highly collimated particle beam and are interrogated in a $\Lambda$-type excitation scheme of optical transitions to long range levels close to the asymptote of the ground state, using the electronically excited state A~$^1\Sigma^+_{\rm u}$ as intermediate one. The transition frequencies  are measured either by comparison with I$_2$ lines or by absolute measurements using a fs-frequency comb. The determined level energies were used together with Feshbach resonances from cold collisions of $^{39}$K and $^{40}$K reported from other authors to fit new ground state potentials. Precise scattering lengths are determined and tests of the validity of the Born-Oppenheimer approximation for the description of cold collisions at this level of precision are performed.
\end{abstract}

\maketitle
%
%
%
%
\section{\label{sec:intro}Introduction}
Progress in experiments with cold, ultracold, quantum degenerate atomic or molecular ensembles starting from laser cooled atoms is steady and fast. An appropriate model of binary collisions is essential for many current and future experiments. The scattering length characterizes the phase shift of the atom pair wave function when one atom slowly propagates through the potential induced by the other atom. This parameter reduces the interatomic potential to an effective phase shift in the limit of zero kinetic energy. It can describe cold collisions, e.g. the dynamics of thermalization or the stability of Bose-Einstein condensates, and is a helpful parameter in the understanding of phase diagrams of diluted ultracold ensembles.

However, a parametrization of the scattering process only by the scattering length is often not sufficient, when tunable resonances are involved that e.g. alter the scattering length \cite{tie93} or allow the formation of quantum degenerate ensembles of dimers \cite{due04} and demand thorough investigations of phase transitions \cite{bar04, gre02}. In many cases, models for ultracold collisions were obtained from resonance  spectroscopy even without deep knowledge of the interatomic potentials at small internuclear separations \cite{tie96, wil08}.

Measuring the binding energy of few loosely bound molecular levels may yield enough information to derive the scattering length \cite{bar05a} or Feshbach resonances, but one relies on long-range interaction parameters calculated from atomic properties. It is also possible to measure the energy position of a larger number of weakly bound molecular levels with molecular spectroscopy, i.e., relative to a deeply bound molecular level \cite{elb99,all03}, or by photoassociation spectroscopy relative to the atomic asymptote \cite{vog00,wan00,van04a}. With such an approach, the determination of the asymptotic energy with respect to deeply bound levels and the long-range interaction coefficients of the potential are reliable if the number of observed levels is high and their binding energy is small. This evaluation can be combined with calculated long-range parameters from dispersion theory since even with a multitude of observed states the values of long range parameters are significantly correlated and thus difficult to determine experimentally in a reliable way. 

If all available data are combined, ground state potential energy curves can be constructed for the whole interatomic separation axis from the repulsive range via the bound one to the asymptote, and they contain the full information of the molecular level structure and can be used to calculate scattering parameters and resonance positions. With the power of the full-potential method being developed in a preceding experiment on sodium \cite{sam00}, the work presented here aims in searching for limits of the potential method. The Born-Oppenheimer approximation (BOA) underlies the potential description for molecules. In this approximation, different isotopes are described by the same potentials but by the use of different reduced masses, hyperfine and Zeeman parameters in the Hamiltonian. Therefore, it is possible to scale the energy levels, Feshbach resonance positions, and scattering lengths from one isotope combination to another. However, this assumes the BOA and any deviation after the corrections for hyperfine structure and mass difference has to be considered a violation of it.

Potassium has not only the two stable bosonic isotopes $^{39}$K and $^{41}$K but also the long-living, naturally occurring $^{40}$K. The large number of accessible isotope combinations makes potassium a good candidate for investigations of BOA violations.  $^{40}$K is of great interest due to its fermionic character \cite{reg03, wil08}. Feshbach resonances have been measured to a high precision for this isotope in several studies \cite{lof02a, reg03b, reg03a, tic04, reg04, gae07} and recently also for $^{39}$K \cite{der07}. Moreover, photoassociation spectroscopy for $^{39}$K  was reported by Wang et al. \cite{wan00}, and a rich molecular data set obtained mostly on $^{39}$K$_2$ is available for both the singlet \cite{ami95} and the triplet ground state  \cite{pas08}. A small number of vibrational levels close to the asymptote could not be observed, which makes the global description by the existing data incomplete and prohibits conclusions about violations of the BOA. 

We report here the observation of levels within the un-observed energy interval in a multi-photon process. In our study, we use long-range levels of the first electronically excited state A~$^1\Sigma^+_{\rm u}$  that we recently investigated \cite{fal06a} and for which we found indications of non-zero corrections to the BOA for levels very close to the asymptote of this state \cite{fal07}. We would like to point out that BOA corrections discussed therein and in this work are not due to hyperfine coupling between different levels, since these perturbations are accounted for in the applied Hamiltonian. Non-adiabatic corrections of the BOA are observed for chemical reactions of light reaction partners are reported for spin-orbit coupling by Li Che et al. \cite{che07a}. Estimates of adiabatic corrections of the BOA, in a similar way like the normal mass effect of atomic energy levels, indicate for the molecular potential energy curves that extended models for cold collisions might be needed at a level of precision of 0.1G or below for Feshbach resonances and molecular levels spanning an energy range of several thousand wave numbers with a precision of 0.001~\wn\ or below as presented here.

The laser-spectroscopic experiments of this study are described in Sec.~\ref{sec:exp}, followed by a detailed analysis of the frequency measurement in Sec.~\ref{sec:meas}. In Sec.~\ref{sec:theo} we introduce our model with parameterized potential energy curves, which allows us to calculate rovibrational levels including hyperfine structure and Zeeman effect, Feshbach resonance positions and scattering lengths. The already existing and newly obtained experimental data are used to derive quantitative results of the potential curves, as discussed in Sec.~\ref{sec:res}. Moreover, we investigate to which precision the BOA is valid by comparing the resulting potentials from fits of different combinations of isotopic data, thus the mass dependence of the interatomic interaction potentials.
%
%
%
%
\section{\label{sec:exp}Experimental Setup}
\begin{figure}[t]
\centerline{\includegraphics[width=9cm]{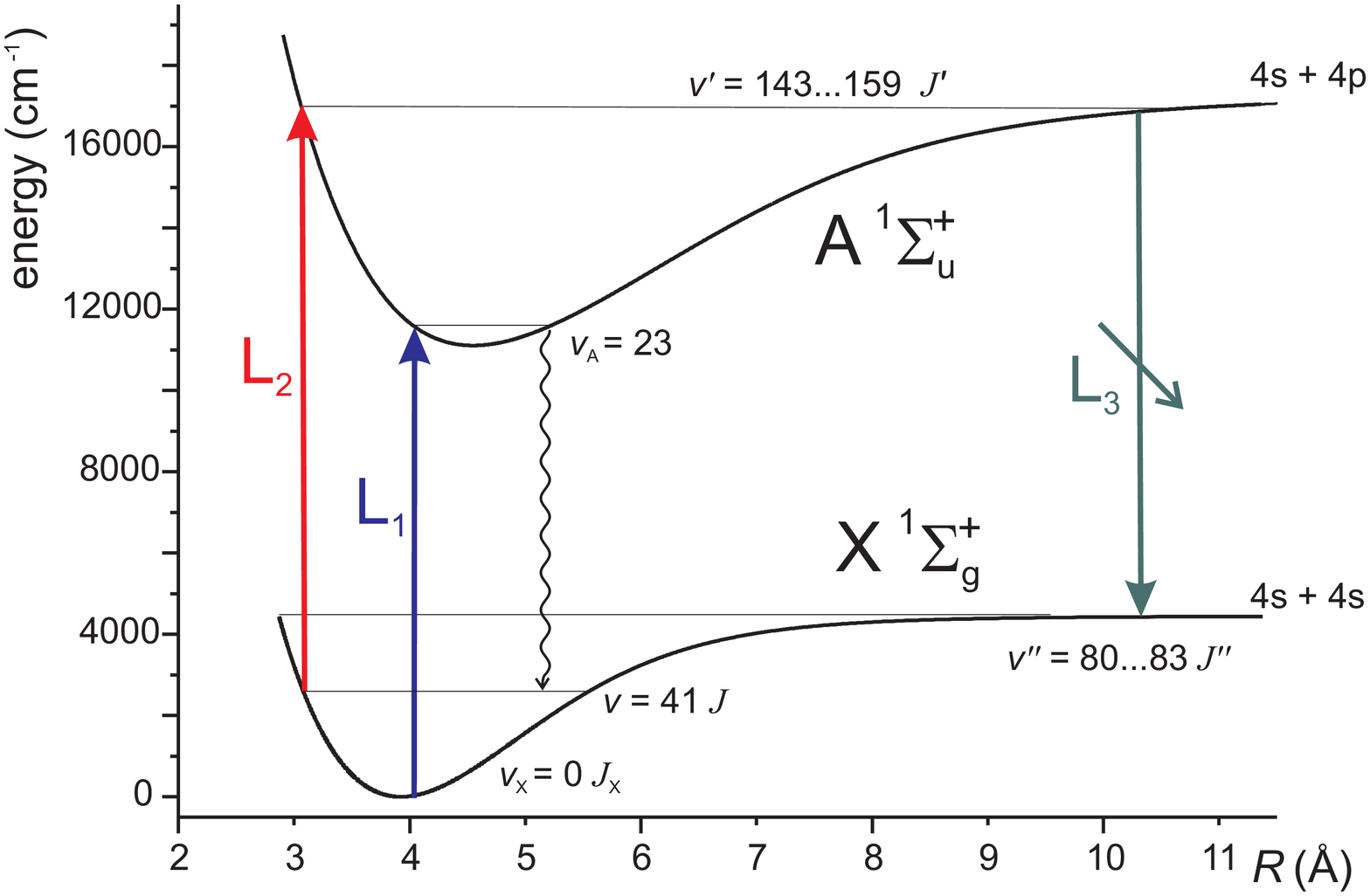}}
		 		  \caption{(Color online) Simplified excitation scheme of K$_2$. Initially, the molecules are in the vibrational ground state of the \Xstate\ state. The laser L$_1$ is applied first and drives Franck-Condon pumping to populate $v=41$, $J$ in the ground  state. Asymptotic levels are observed as dark resonances induced by laser L$_2$ (on resonance) and laser L$_3$ (tuned).}
		 		  \label{fig:pump}
\end{figure}

The setup, which we use to create a beam of K$_2$ molecules, was described in some detail in a previous publication \cite{lis01a} and was well characterized for precision measurements \cite{fal06,she07} and Ramsey-Bord\'e interferometry with molecules \cite{lis00}. In short, potassium is heated in a furnace to about 700~K. Through a 200~$\mu$m nozzle, potassium vapour expands into the vacuum and forms an atomic and molecular beam with a mean particle velocity of about 1000~m/s.

Mainly, the vibrational ground state $v_{\rm X}=0$ of the electronic state \Xstate\ is populated. Due to very unfavourable Franck-Condon factors, no direct coupling scheme in a $\Lambda$-configuration to asymptotic levels of the electronic ground state can be driven starting from this level. Therefore, we employ optical pumping in a first interaction zone of the molecular beam. Laser L$_1$ transfers a significant amount of population to higher vibrational levels \cite{sam00,fal06a,fal07} of the ground state. This pumping step is achieved by an extended cavity diode laser (ECDL) at 792~nm in Littrow configuration. The light is transfered with a fibre to the vacuum system, where it excites a rotational line P($J_X$) or R($J_X$) in the 23-0 band of the A-X system of $^{39}$K$_2$ (see Fig.~\ref{fig:pump}). Spontaneous decay populates with about 15\% probability the vibrational level $v=41$, $J=J_X$, $J_X-2$ or $J_X$, $J_X+2$, respectively. 

Molecules in excited vibrational levels like $v=41$ have a very long lifetime and will reach the second interaction zone about 30 cm downstream the molecular beam, where they interact with two co-propagating laser beams L$_2$ and L$_3$, which are also delivered by fibres. Laser fields and the molecular beam interact under a 90$^{\circ}$ angle. The laser beams have diameters of a about 2~mm and are superimposed on a dichroic mirror. These two fields will establish the $\Lambda$-type coupling scheme connecting the ground state vibrational levels $v$ and $v''$ via the vibrational level $v'$ in the excited state A~$^1\Sigma^+_{\rm u}$. The energetic positions of the vibrational levels in the A~state were investigated previously and are tabulated in \cite{fal06a}. 

Laser induced fluorescence can be observed from the second interaction zone. The laser L$_2$ is held on resonance for one of the molecular transitions $(v',J'=J\pm 1) \leftarrow (v,J)$ at about 710~nm and induces fluorescence, mainly at 775~nm. L$_2$ is a titanium:sapphire laser pumped by a 10~W pump laser at 532~nm. It is frequency stabilized by an offset-lock to an iodine-stabilized HeNe laser. For this purpose, L$_2$ is locked to a tunable transfer cavity, which is itself stabilized to the HeNe laser. An acousto-optical modulator (AOM) in the beam path of the HeNe laser allows a shift of one transverse mode of the transfer cavity to any arbitrary frequency of L$_2$.

Laser L$_3$ is a second Ti:sapphire laser with a 5~W pump laser. It is operated around 775~nm  and tuned. If L$_3$ is resonant with a transition $(v',J') \rightarrow (v'',J''=J'\pm 1)$, a dark resonance or electromagnetically induced transparency is induced and the fluorescence from the excited state is reduced. Unfortunately, the fluorescence spectrally overlaps with stray light of laser L$_3$. Thus, discrimination by optical filters is not possible and, therefore, direct observation of the dark resonances was not successful. We could overcome this problem of detecting the resonances in presence of laser stray light by modulating laser L$_1$ with a chopper and filtering the fluorescence signal with a lock-in amplifier driven at the chopper frequency.

The signal-to-noise ratio was still small, so the laser L$_3$ had to be tuned over the dark resonance with high resolution and very good reproducibility to be able to average several independent measurements for improving the signal-to-noise ratio. Additionally, its frequency must be precisely known (as for laser L$_2$) to calculate the energy difference between different asymptotic levels $v''$, $J''$ and the levels $v$, $J$. We met these requirements by means of a two step offset stabilization of L$_3$ with respect to an iodine stabilized diode laser (see below) of well known frequency. The frequency of laser L$_2$ was measured in a later experimental run by a fs-frequency comb referenced to a GPS-disciplined quartz oscillator. A similar scheme as for L$_3$ was not possible for L$_2$ because no I$_2$ lines are known with the required accuracy in this spectral interval.

\begin{figure}[t]
\centerline{\includegraphics[width=9cm]{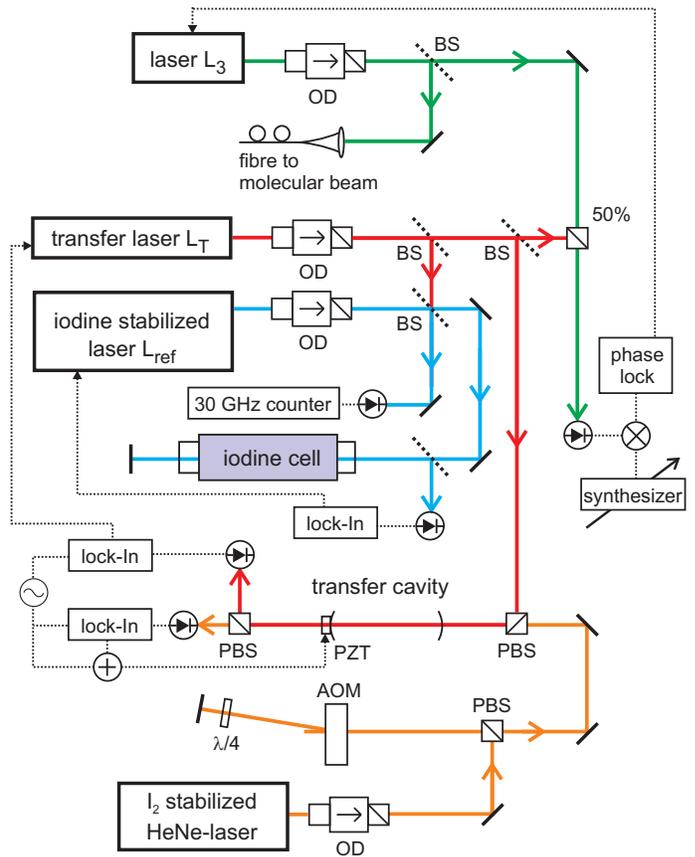}}
		 		  \caption{(color online) Laser system used for frequency measurement and tuning of L$_3$. OD: optical diode; BS: beam splitter;  PBS: polarizing beam splitter; PZT: piezo-electric actuator;  $\lambda$/4: $\lambda$/4 retarder; AOM: acousto-optical modulator.}
		 		  \label{fig:laser}
\end{figure}

Figure \ref{fig:laser} shows the offset-stabilization scheme of L$_3$. Iodine lines of the (0-14) and (1-14) bands of the B-X system with an uncertainty of about 1~MHz were used as references. Though the I$_2$ spectrum is very dense, frequency intervals of tens of GHz had to be bridged to reach the resonances of L$_3$. For this purpose we stabilized an ECDL L$_{\rm ref}$ by saturated absorption spectroscopy to the required transitions. We used 70~cm long iodine cells heated to 740~K with a side arm temperature of 295~K. In principle, L$_3$ could have been stabilized and tuned by stabilizing the beat note of lasers L$_3$ and L$_{\rm ref}$ to an appropriate tunable high frequency source. However, the latter was not available. Instead, we introduced an ECDL transfer laser L$_{\rm T}$, which is closer in frequency than 1~GHz to L$_3$. L$_{\rm T}$ is frequency stabilized like L$_2$ via another transfer cavity relative to the I$_2$-stabilized HeNe laser. The beat note between L$_{\rm ref}$ and L$_{\rm T}$ was observed with a fast photodiode, counted by a rf-counter, and recorded for later corrections. The beat note between L$_3$ and L$_{\rm T}$ was mixed with a tunable synthesizer and the mixing product was stabilized by means of a digital divider and a phase-locked-loop. In this setup we transfer the long-term stability and accuracy from the I$_2$ lines to the laser L$_3$ and achieve reliable tuning of the laser. This laser tuning scheme has already proved to be successful for high precision frequency measurement with potassium atoms \cite{fal06}.

%
%
%
%
\section{\label{sec:meas} Measurements and Uncertainties}
%
%
%
%
\subsection{\label{sec:data} Data Recording}
\begin{figure*}[t]
\centerline{\includegraphics[width=18cm]{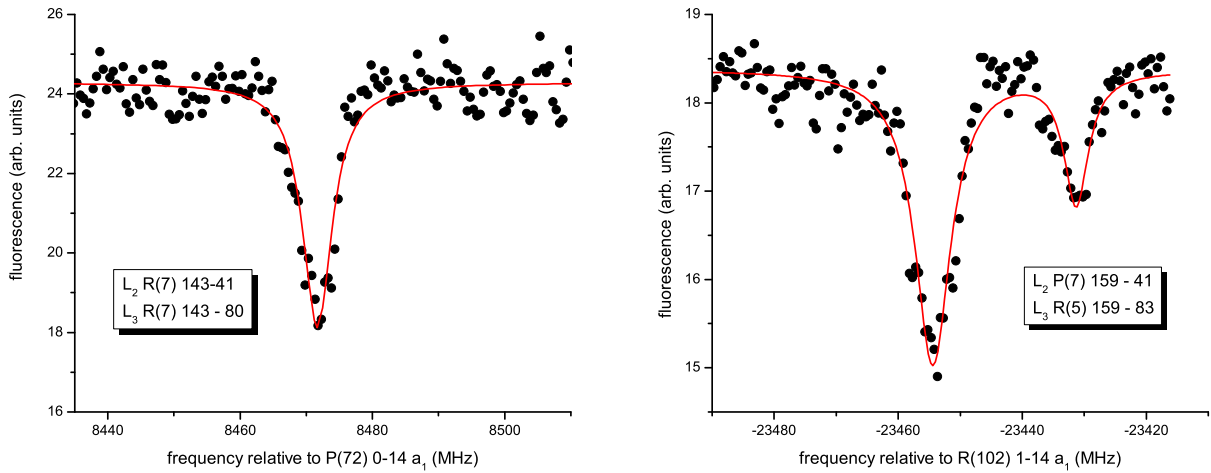}}
		 		  \caption{(Color online) Spectra of dark resonances observed in two $\Lambda$-configurations. Left-hand side: coupling between $(v=41,J=7)-(v'=143,J'=8)-(v''=80,J''=7)$. Right-hand side: case $(v=41,J=7)-(v'=159,J'=6)- (v''=83,J''=5)$, two dips are visible, which belong to the small hyperfine splitting of asymptotic levels in the state~X. The hyperfine structure stems from coupling to the \astate\ state and is so small in potassium that it is only resolved for $v''=82$ and higher. The frequency scale is given with respect to iodine lines}
\label{fig:spec}
\end{figure*}
The aim of the present spectroscopic measurements was to determine energy differences between asymptotic levels of $^{39}$K$_2$ and differences of asymptotic levels to levels in the well bound region of the potential. For this purpose, the investigations were divided into two steps: the spectroscopy of dark resonances with frequency determination of L$_3$ while L$_2$ was held on a molecular resonance. Later we measured with a fs-frequency comb the absolute frequency of those transitions that turned out to be successful for the spectroscopy of the dark resonances. Systematic frequency shifts between the two sets of measurements have to be avoided. The main source of uncertainty is here the first-order Doppler effect introduced by an angle different from 90$^{\circ}$ between the laser beams and the molecular beam (see Sec.~\ref{sec:err}).

We started searching for ground state resonances in the region of $v''=80$ where the measurements from classical spectroscopy \cite{ami95} stopped and left an energy gap of about 1.8~cm$^{-1}$ to the asymptote 4s + 4s. Due to selection rules, for each level ($v',J'$) two dark resonances with $J''=J'\pm 1$ are expected for each vibrational level $v''$. In most cases, it is practically impossible to observe more than one vibrational level $v''$ from a given level $v'$ because the Franck-Condon overlap drops off dramatically with the rapidly growing classical outer turning point of the vibration wave function when $v''$ is increased by one unit.

In Fig.~\ref{fig:spec}, observed dark resonances for two levels $(v'',J'')$ are shown. A single resonance appears (left-hand side) for the coupling between $(v=41,J=7)-(v'=143,J'=8)-(v''=80,J''=7)$ when L$_3$ is tuned. On the right-hand side, two resonances are visible that are due to hyperfine splitting of the level $(v'' = 83, J''=5)$. The structure in the asymptotic levels of the state \Xstate\ appears by hyperfine coupling to the \astate\ state and is small for K$_2$ compared to Na$_2$ \cite{elb99} or Cs$_2$ \cite{wei86}. We could only resolve the splitting for vibrational levels $v''=82$ and higher. 

The hyperfine component at lower frequency in the right-hand part of Fig.~\ref{fig:spec} corresponds to a total nuclear spin $I = 3$, the high frequency one to $I=1$. Since $J''$ is odd, $I=0,2$ do not appear for symmetry reasons \cite{lis00}. The intensity ratio of the two observed dips in the spectrum is very close to the expected ratio of 7/3.

We have have observed in total 14 asymptotic levels of which the least bound one is only 0.182~cm$^{-1}$ below the lowest hyperfine dissociation threshold and has an outer turning point of about 21 \AA. The assignment of quantum numbers to all levels $(v'',J'')$ is based on selection rules for the rotation, and counting of vibrational levels provided by Amiot \cite{ami95,amiPC}, which was later confirmed by our analysis (Sec.~\ref{sec:res}). Frequencies, uncertainties and quantum numbers of all resonances measured by tuning laser L$_3$ are given in Tab.~\ref{tab:freq}.
\begin{table}
  \centering
\caption{\label{tab:freq} Quantum numbers and frequencies $\nu$ of transitions $(v'-v'')$ observed by tuning L$_3$. If the hyperfine splitting could be resolved, the quantum number of the total nuclear spin $I$ is given. The number in brackets behind the frequency $\nu$ is the uncertainty in units of the last digit. In the last column, the iodine reference lines are listed~\cite{kno04}.}
\begin{tabular}{ c c l l }\hline
$v'-v''$ & Line &  \makebox[2.5cm][c]{$\nu$ in cm$^{-1}$} & \makebox[2cm][c]{Iodine line}\\
\hline
\hline
143-80 & R(5)  & \makebox[2.5cm][r]{12\:813.819\:53(7)} & P(72) 0-14 a$_1$\\
143-80 & R(7)  & \makebox[2.5cm][r]{12\:813.929\:19(7)} & P(72) 0-14 a$_1$\\
143-80 & R(9)  & \makebox[2.5cm][r]{12\:814.062\:36(7)} & P(72) 0-14 a$_1$\\
143-80 & P(11) & \makebox[2.5cm][r]{12\:813.761\:65(7)} & P(72) 0-14 a$_1$\\
147-81 & R(1)  & \makebox[2.5cm][r]{12\:843.431\:73(7)} & R(42) 0-14 a$_{15}$\\
147-81 & P(3)  & \makebox[2.5cm][r]{12\:843.371\:59(7)} & R(42) 0-14 a$_{15}$\\
147-81 & R(5)  & \makebox[2.5cm][r]{12\:843.587\:27(7)} & R(42) 0-14 a$_{15}$\\
147-81 & P(7)  & \makebox[2.5cm][r]{12\:843.433\:71(7)} & R(42) 0-14 a$_{15}$\\
147-81 & R(9)  & \makebox[2.5cm][r]{12\:843.845\:56(7)} & R(42) 0-14 a$_{15}$\\
147-81 & P(11) & \makebox[2.5cm][r]{12\:843.605\:99(7)} & R(42) 0-14 a$_{15}$\\
159-82 & R(9) $I=3$ & \makebox[2.5cm][r]{12\:910.763\:55(8)} & R(102) 1-14 a$_1$\\
159-82 & R(9) $I=1$ & \makebox[2.5cm][r]{12\:910.763\:84(9)} & R(102) 1-14 a$_1$\\
159-82 & P(11) $I=3$& \makebox[2.5cm][r]{12\:910.587\:20(8)} & R(102) 1-14 a$_1$\\
159-82 & P(11) $I=1$& \makebox[2.5cm][r]{12\:910.587\:52(9)} & R(102) 1-14 a$_1$\\
159-82 & P(7) $I=3$ & \makebox[2.5cm][r]{12\:910.486\:21(8)} & R(102) 1-14 a$_1$\\
159-82 & P(7) $I=1$ & \makebox[2.5cm][r]{12\:910.486\:41(9)} & R(102) 1-14 a$_1$\\
159-83 & R(5) $I=3$ & \makebox[2.5cm][r]{12\:910.100\:56(8)} & R(102) 1-14 a$_1$\\
159-83 & R(5) $I=1$ & \makebox[2.5cm][r]{12\:910.101\:33(8)} & R(102) 1-14 a$_1$\\
\hline
\end{tabular}
\end{table}

After the determination of the transition frequencies observed by L$_3$ we have measured the frequency of L$_2$, when exciting the upper level $(v',J')$, with the help of a commercial fs-fibre laser frequency comb, which was referenced to a GPS-disciplined quartz oscillator. To avoid systematic frequency shifts from first-order Doppler effect, laser L$_2$ was aligned with the molecular beam by the same procedure (Sec.~\ref{sec:err}) as before and was stabilized with the offset-lock to the iodine-stabilized HeNe laser to stay resonant with the molecular transition under investigation.

Each frequency setting of laser L$_2$ was typically measured for 80~s with a gate time of the counters of 1~s. The observed scatter of the measured frequencies was below 50~kHz. The mode number of the frequency comb was unambiguously determined by measuring  the wavelength of L$_2$ with a commercial wavelength meter corresponding to 30~MHz accuracy. The wavemeter was calibrated by measuring a laser stabilized to the Mg intercombination line \cite{fri08} or to the iodine stabilized HeNe laser used earlier for L$_3$.

The frequencies, uncertainties and quantum numbers of all transitions pumped by L$_2$ are given in Tab.~\ref{tab:freq2}. From the frequencies given in Tab.~\ref{tab:freq} and Tab.~\ref{tab:freq2}, various energy intervals within the manifold of rovibrational levels of the ground state \Xstate\ can be calculated that were used for the fits yielding potentials with high precision (Sec.~\ref{sec:res}).
\begin{table}
  \centering
\caption{\label{tab:freq2} Quantum numbers and frequencies $\nu$ of the transitions pumped by L$_2$. The number in brackets behind the frequency $\nu$ is the uncertainty in units of the last digit.}
\begin{tabular}{ c c l }\hline
$v'-v$ & Line &  \makebox[2.5cm][c]{$\nu$ in cm$^{-1}$} \\
\hline
\hline
143-41 & P(7)  & \makebox[2.5cm][r]{14\:057.684\:43(7)} \\
143-41 & R(7)  & \makebox[2.5cm][r]{14\:057.984\:44(7)} \\
143-41 & P(11) & \makebox[2.5cm][r]{14\:055.057\:67(7)} \\
147-41 & P(3)  & \makebox[2.5cm][r]{14\:090.628\:40(7)} \\
147-41 & P(7)  & \makebox[2.5cm][r]{14\:089.036\:59(7)} \\
147-41 & R(9)  & \makebox[2.5cm][r]{14\:088.166\:78(7)} \\
159-41 & R(9)  & \makebox[2.5cm][r]{14\:155.964\:91(7)} \\
159-41 & P(7)  & \makebox[2.5cm][r]{14\:157.018\:56(7)} \\
\hline
\end{tabular}
\end{table}
%
%
%

%
%
\subsection{\label{sec:err} Uncertainty Considerations}
Several aspects have to be considered to determine the uncertainty of the measured transition frequencies. The largest contribution is expected from the uncertainty of the adjustment of the laser beams L$_2$ and L$_3$ under a 90$^{\circ}$ angle to the molecular beam. Any deviation from perpendicular adjustment will introduce systematic shifts due to first-order Doppler effect.

We always followed the same alignment procedure in each series of measurements. Laser L$_2$ was tuned to the potassium D1 line. The atomic line was recorded at low laser intensity to avoid saturation broadening and in two distinct configurations: with a single pass laser beam and with the laser beam retro-reflected. In the second configuration, the beam was reflected back by a mirror about 2~m from the fibre output. The parallelism of the two counter-propagating beams was checked with an aperture in front of the fibre coupler. 

Then the observed line width in both configurations was compared. A deviation from 90$^{\circ}$ in the crossing angle of laser and particle beam will show up in a broadening of the line profile in the two-beam configuration. Systematic variation of the angle allowed to determine a minimum of the line broadening with an uncertainty of about 0.5~mrad, which corresponds to an uncertainty in the frequency of L$_2$ of 0.7~MHz due to the Doppler shift.

For the alignment of L$_3$ with respect to L$_2$, an aperture in front of the mirror retro-reflecting L$_2$ was used. Again, an uncertainty of 0.5~mrad in the angle was achieved, which leads to an additional possible Doppler shift of 0.6~MHz after alignment of L$_3$ relative to L$_2$.

The second important contribution to the uncertainty is the reliable determination of the resonance frequency for laser L$_2$ because the frequency measurements of L$_2$ and L$_3$ were separated in time and the dark resonance is observed at the two-photon resonance. We found that we can determine the line centre to within 1.8~MHz by tuning the AOM frequency of the offset-lock of L$_2$ to the HeNe laser. This uncertainty is much larger than the observed drift of the applied offset-lock during this or previous experiments \cite{fal06}.

Further contributions stem from the determination of the line centre of the dark resonance (typically 0.3~MHz, depending on the signal-to-noise ratio), the frequency uncertainty of the iodine lines used for calibration (0.1~MHz, for the 1-14 band 1.3~MHz), and the possible uncertainties in the lock of the spectroscopy laser to the transfer laser, synchronisation of the 
counter traces with the spectra etc. (in total 0.3~MHz). The uncertainty of the frequency measurement with the fs-comb was estimated to 0.2~MHz. Other factors like Zeeman effect or second-order Doppler effect are negligible. The respective uncertainty contributions for the measured lines were added quadratically. A compilation of all contributions is listed in Tab.~\ref{tab:err}.

\begin{table}
  \centering
\caption{\label{tab:err} Summary of the different contributions to the uncertainty budget.}
\begin{tabular}{ l c }\hline
source          &  uncertainty\\
\hline
\hline
Doppler effect L$_2$ & 0.7~MHz\\
Doppler effect L$_2$ rel. to L$_3$ & 0.6~MHz\\
centre of excitation L$_2$ & 1.8~MHz\\
centre of dark resonance L$_3$& 0.3~MHz\\
offset lock L$_3$, counter& 0.3~MHz\\
iodine lines & 0.1~MHz / 1.3~MHz\\
fs-comb& 0.2~MHz\\
\hline
\end{tabular}
\end{table}

%
%
%
%
\section{\label{sec:theo} Theoretical description}
To cover the full range of experimental data from Fourier transform spectroscopy over molecular beam spectroscopy to Feshbach resonances by a theoretical model, only a coupled channel analysis is adequate, which includes both molecular states \Xstate\ and \astate, the hyperfine coupling and the Zeeman interaction, to compute the rotational, vibrational and hyperfine structure of the bound states and the scattering states. Such a theoretical approach has been described in several papers, e.g. \cite{pas07}, so we invite the reader to follow our construction of the theoretical model that is able to reproduce all existing measurements within their reported experimental uncertainty.

The full Hamiltonian, see for example Refs.~\cite{mie00,lau02,pas07}, for a pair of atoms $A$ and $B$ with electron and nuclear spin $s$ and $i$, respectively, can be written in the form: 

\begin{eqnarray}
\label{eq:ham}
 H  &=& T_n+U_{\rm X}(R)P_{\rm X} + U_{\rm a}(R)(1-P_{\rm X})\nonumber\\
 & &+a_A(R)\vec{s}_A\vec{i}_A+a_B(R)\vec{s}_B\vec{i}_B\nonumber\\
 & &+(g_{sA}s_{zA}-g_{iA}i_{zA})\mu_{\rm Bohr} B_z+(g_{sB}s_{zB}-g_{iB}i_{zB})\mu_{\rm Bohr} B_z\nonumber\\
 & &+\frac{2}{3} \lambda(R)(3S_Z^2-S^2) 
\end{eqnarray}

The first line shows the kinetic energy $T_n$ and the potential energy $U_{\rm X}$ and $U_{\rm a}$ for the motion of the atoms, where $P_{\rm X}$ and $1-P_{\rm X}$ are projection operators on the uncoupled states X and a, respectively. The second line shows the hyperfine interaction, determined mainly by the Fermi contact term, and the functional dependence with $R$ for the hyperfine parameters indicates that distortions of the considered atom from the second one should be expected. But the coupling of nucleus $A$ with the electron spin of atom $B$ is neglected, and also the interaction with the nuclear quadrupole moment coming into play for more deeply bound levels. The third line gives the Zeeman energy from the electron spin and the nuclear spin by an external homogeneous magnetic field $B$ in $z$ direction. The last line contains the spin-spin interaction represented by the total molecular spin $S$ and its direction to the molecule fixed axis Z. The parameter $\lambda$ is  a function of $R$, mainly $1/R^3$ as dipole-dipole interaction, but it contains also contributions from second order spin-orbit interactions. The final analysis showed that such later contribution is insignificant within the achieved experimental accuracy in the case of potassium.

We take the magnetic hyperfine parameters and the electronic and nuclear $g$-factors for the atomic ground state of the potassium isotopes from the report in \cite{ari77}, because the hyperfine structure was only seen for asymptotic vibrational levels and Feshbach resonances, thus any dependence of the hyperfine interaction on the internuclear separation by chemical bonds are probably very small and not identified in our analysis. The atomic masses are taken from the recent tables by G.~Audi et al. \cite{aud03}.

The calculations incorporate Born-Oppenheimer potentials according to Hund's coupling case b, because the total electronic spin is taken as good quantum number for U$_{\rm X}$ and U$_{\rm a}$.

Each representation of the potentials for the two states with the common atomic asymptote 4s + 4s is split into three regions on the internuclear separation axis $R$: the repulsive wall ($R < R_\mathrm{inn}$), the asymptotic region ($R > R_\mathrm{out}$),
and the intermediate region in between. The analytic form of the potentials in the intermediate range is described by a finite power expansion with a nonlinear variable function $\xi$ of internuclear separation $R$:

\begin{equation}
\label{eq:xv}
\xi(R)=\frac{R - R_m}{R + b\,R_m}
\end{equation}
\begin{equation}
\label{eq:uanal}
U_{\mathrm {IR}}(R)=\sum_{i=0}^{n}a_i\,\xi(R)^i,
\end{equation}

\noindent where the $a_i$ are fitting parameters and $b$  and $R_m$ are chosen such that only few parameters $a_i$ are needed for describing the steep slope at the short internuclear separation side and the much smaller slope at the large $R$ side by the analytic form of Eq.~(\ref{eq:uanal}), $R_m$ is normally close to the value of the equilibrium separation. The potential is extrapolated for $R < R_{\rm inn}$ with:

\begin{equation}
\label{eq:rep}
  U_{\mathrm {SR}}(R)= A + B/R^{N_s}
\end{equation}

\noindent by adjusting the $A$ and $B$ parameters to get a continuous transition at $R_{\rm inn}$; 
the final fitting uses $N_s$ equal to 12 and 6 for \Xstate\ and \astate\ states, respectively, as adequate exponents.

For large internuclear distances ($R > R_{\rm out}$)
we adopted the standard long range form of molecular potentials:

\begin{equation}
\label{eq:lrexp}
  U_{\mathrm {LR}}(R)=U_{\infty}-C_6/R^6-C_8/R^8-C_{10}/R^{10}\pm E_{\mathrm{exch}},
\end{equation}

\noindent where the exchange contribution is given by

\begin{equation}
\label{eq:exch}
E_{\mathrm{exch}}=A_{\mathrm{ex}} R^\gamma \exp(-\beta R) 
\end{equation}
and U$_{\infty}$ set to zero for fixing the energy reference of the total potential scheme.

The data on hand include three different isotopomers, namely $^{39}$K$_2$, $^{40}$K$_2$, and $^{39}$K$^{41}$K. Thus it is possible to check the validity of the Born-Oppenheimer approximation or the so called mass scaling of cold collision properties between different isotopomers. For this purpose the mechanical Schr\"odinger equation for the rovibrational motion should be reformulated to allow for adiabatic and non-adiabatic corrections to the Born-Oppenheimer approximation. The most often used form (see e.g. \cite{kno04}) of the radial equation for a rotational state $J$ is:

\begin{eqnarray}
{\bf H_{\rm eff} }& = & 
 -\frac{\hbar^2}{2\mu}\frac{\partial}{\partial R}[1+\beta(R)]\frac{\partial}{\partial R}
+U(R)+U_{\rm ad}(R) \nonumber\\
 & &+\frac{\hbar^2\cdot [1+\alpha(R)]\cdot J(J+1)}{2\mu R^2}
\label{eq:BOC}  
\end{eqnarray}

For a homonuclear molecule the corrections $\alpha$, $\beta$ and $U_{\rm ad}$ depend on the ratio $m_e$/$\mu$ of the electron mass $m_e$ and the reduced mass $\mu$ of the molecule. Because only two of these corrections can be determined simultaneously \cite{tie94}, we decided to neglect $\beta$ in our approach. In this way $\alpha(R)$ becomes an effective non-adiabatic correction, not only the rotational correction. Additionally, the Schr\"odinger equations gets the proper form for direct application of the conventional Numerov procedure to calculate eigenvalues and eigenfunctions. The adiabatic correction $U_{\rm ad}(R)$ can only be determined, if data of different isotopic species of the molecule are included. Both $\alpha$ and $U_{\rm ad}$ are functions of the internuclear separation $R$ and will be represented by  truncated power series in $\xi$ as the potential curves:

\begin{equation}
\alpha(R) =\frac{\mu_{\rm ref}}{\mu}\cdot \frac{2R_m}{R+R_m} \sum_{i=0,1,...} \alpha_{i} \cdot \xi^i 
\label{eq:BOC1}
\end{equation}
\begin{equation}
U_{\rm ad}(R) = \left( 1-\frac{\mu_{\rm ref}}{\mu} \right) \cdot \left( \frac{2R_m}{R+R_m} \right)^6 
\sum_{i=0,1,...} v_i \cdot \xi^i    
\label{eq:BOC2}
\end{equation}
 
with $\mu_{\rm ref}$ being the reduced mass of the selected reference isotope, here $^{39}$K$_2$. These equations consider the asymptotic behaviour of the corrections, which should vanish for $R \rightarrow \infty$ because the intermolecular motion becomes infinitely slow, and for the adiabatic correction the proper exponent, namely 6, to keep the lowest order of the long range behavior as in Eq.~(\ref{eq:lrexp}). The adiabatic corrections will be added to the potentials in Eq. (\ref{eq:ham}) and the operator T$_n$ will be conventionally split into the one dimensional kinetic energy of vibration and the rotational energy where $\alpha(R)$ can be added according to Eq. (\ref{eq:BOC}). With this we have now the complete physical model for calculating the necessary eigenenergies for the spectral data and the scattering properties to derive the resonances.

The evaluation for the free parameters in the model is performed in a self-consistent iteration loop, in which the fit of the Born-Oppenheimer potentials, that describe the spectroscopic data with the help of the Numerov procedure for solving the Schr\"odinger equation as given for the states \Xstate\ and \astate\ separately by Eq.(\ref{eq:BOC}), is alternated with the coupled channels calculations for describing the Feshbach resonances and photoassociation data. The coupled channels calculations are also used for deriving the hyperfine corrections to the observed spectral lines and thus constructing the hyperfine free spectroscopic data, which are the inputs to the potential fits. The potentials are obtained simultaneously for both states, because we have data which represent the difference between the singlet and triplet manifold of rovibrational levels and both states have a common asymptote with a common long range function as given by Eq.~(\ref{eq:lrexp}). This is a very important constraint for both potential functions, often neglected within the evaluation of sets of potential functions.
%
%
%
%
\section{\label{sec:res} Evaluation and Results}
The most recent analysis of the two ground states of potassium is from our own group~\cite{pas08}. This work used the spectroscopic data from Fourier transform spectroscopy on the state \Xstate\ from \cite{ami95} and own measurements for this state and the state \astate, and furthermore, results from Feshbach spectroscopy by \cite{reg03b,reg04,gae07,der07} and photoassociation spectroscopy by \cite{wan00}. With the present beam experiment, the remaining energy gap between the range covered by conventional spectroscopy results and those from the Feshbach resonance studies of about 0.9~\wn\ and 1.7~\wn\ for the states \Xstate\ and \astate, respectively, could be closed for the state \Xstate, but no new measurements were obtained for the state \astate. However, the observed hyperfine structure of the highest vibrational levels in \Xstate, see Fig. \ref{fig:spec}, is a clear signature of the coupling between the singlet and the triplet state, and introduces new information. 

To learn how much could be gained by these additional pieces of information, we calculated the term energies of the newly observed level with the formerly published potential curves \cite{pas08}. All new measurements showed significant deviations to the simulation results. First, the measurements contain very precisely determined energy intervals between $v=41$ and levels from $v''=80$ on, here the deviations were about twenty times the experimental uncertainty and still about two times the expected accuracy predicted from the uncertainty of the Fourier transform data. Second, the differences between pairs of asymptotic levels, say for example $v''=80$ to 83 etc., show deviations in the order of twenty to forty times the experimental uncertainty. Thus the new data give important information for connecting the set of deeply bound levels with the asymptotic ones derived from Feshbach resonances.

We started the full routine of self consistent field approach by iterating between potential and asymptotic fit of spectroscopic data and of  Feshbach resonances, respectively. For these calculations we reduced the weight for the large vibrational intervals ($v=41$ to $v''=80$ and above) by selecting as experimental uncertainty only 0.001~\wn\ instead of the actual value as derivable from tables \ref{tab:freq}, \ref{tab:freq2} to be between 0.0001 and 0.0002~\wn. Otherwise, those high weights might locally distort the potential fit. We already mention here, that the final fit result reproduces these data within the original experimental uncertainty with the exception of two cases where the deviations are two times the uncertainty.  

Because we are interested to use the information on the Feshbach resonances up to their experimental uncertainty, we checked first, if thermal averaging is needed to describe the observed resonance position in the ultracold ensemble at the specific conditions from the different laboratories. We followed the procedure described by Ticknor et al. \cite{tic04} and integrated the energy dependent cross sections or rates of the assumed elastic collision over a Boltzmann distribution, because the experimental conditions were sufficiently far from the regime of quantum degeneracy. Most profiles plotted linearly as function of magnetic field show an asymmetric form, but the peak positions deviated very little from those expected when using as single kinetic energy the value derived from the reported temperature. Thus thermal averaging during the fit procedure was unnecessary. We would like to mention, that the functional form of the cross section with energy given in Fig.~1b in \cite{tic04} for s-wave scattering and rationalized in that paper is not always so simple. We found cases where also peak structures appear, namely if the resonance peak is located at an energy before the unitarity limit is approached. These cases lead to more asymmetry in the resonance profiles as functions of magnetic field as in the cases of p-wave resonances, but we did not find for s-waves such severe examples as given in Fig.~2a of \cite{tic04} for p-waves. We should remind the reader that we simply assume that the resonance position can be derived from an elastic two-body collision, whereas in the experiment often trap loss is observed, and thus a three body process is involved. Deviations have been studied for the case of Rb by Smirne et al. \cite{smi07}. Extending our analysis in this direction is far beyond our present capability, developments of new theoretical modeling approaches, however, would be desirable.

\begin{table}
\caption{Parameters of the analytic representation of the \Xstate\ state potential without any Born-Oppenheimer correction. The energy reference is the dissociation asymptote. Parameters with $^\ast$ are set for continuous extrapolation of the potential. }
\label{tab:potx1}
\begin{tabular*}{0.8\columnwidth}{@{\extracolsep{\fill}}|lr|}
\hline
   \multicolumn{2}{|c|}{$R < R_\mathrm{inn}=$ 2.870 \AA}    \\
\hline
   $A^\ast$ & $-0.262878738\times 10^{4}$ \wn \\
   $B^\ast$ & $0.8129033720\times 10^{9}$  \wn \AA $^{12}$ \\
   $N_s$    & $12$\\
\hline
   \multicolumn{2}{|c|}{$R_\mathrm{inn} \leq R \leq R_\mathrm{out}=$ 12.000 \AA} \\
\hline
    $b$ &   $-0.40$ \\
    $R_\mathrm{m}$ & 3.92436437 \AA  \\
    $a_{0}^\ast$ &  $-$4450.899108 \wn\\
    $a_{1}$ &  0.027435192082021$$ \wn\\
    $a_{2}$ &  $0.13671215240591\times 10^{5}$ \wn\\
    $a_{3}$ &  $0.10750901039993\times 10^{5}$ \wn\\
    $a_{4}$ & $-0.20933147904789\times 10^{4}$ \wn\\
    $a_{5}$ & $-0.19385880603136\times 10^{5}$ \wn\\
    $a_{6}$ & $-0.49208904259548\times 10^{5}$ \wn\\
    $a_{7}$ & $ 0.11026640034823\times 10^{6}$ \wn\\
    $a_{8}$ & $ 0.72867340031285\times 10^{6}$ \wn\\
    $a_{9}$ & $-0.29310679230619\times 10^{7}$ \wn\\
   $a_{10}$ & $-0.12407070105941\times 10^{8}$ \wn\\
   $a_{11}$ & $ 0.40333947204823\times 10^{8}$ \wn\\
   $a_{12}$ & $ 0.13229848870507\times 10^{9}$ \wn\\
   $a_{13}$ & $-0.37617673800749\times 10^{9}$ \wn\\
   $a_{14}$ & $-0.95250413278553\times 10^{9}$ \wn\\
   $a_{15}$ & $ 0.24655585743928\times 10^{10}$ \wn\\
   $a_{16}$ & $ 0.47848257694561\times 10^{10}$ \wn\\
   $a_{17}$ & $-0.11582132110109\times 10^{11}$ \wn\\
   $a_{18}$ & $-0.17022518297748\times 10^{11}$ \wn\\
   $a_{19}$ & $ 0.39469335034300\times 10^{11}$ \wn\\
   $a_{20}$ & $ 0.43141949844175\times 10^{11}$ \wn\\
   $a_{21}$ & $-0.97616955325590\times 10^{11}$ \wn\\
   $a_{22}$ & $-0.77417530686085\times 10^{11}$ \wn\\
   $a_{23}$ & $ 0.17314133615815\times 10^{12}$ \wn\\
   $a_{24}$ & $ 0.96118849114885\times 10^{11}$ \wn\\
   $a_{25}$ & $-0.21425463041524\times 10^{12}$ \wn\\
   $a_{26}$ & $-0.78513081753454\times 10^{11}$ \wn\\
   $a_{27}$ & $ 0.17539493131261\times 10^{12}$ \wn\\
   $a_{28}$ & $ 0.37939637010974\times 10^{11}$ \wn\\
   $a_{29}$ & $-0.85271868689619\times 10^{11}$ \wn\\
   $a_{30}$ & $-0.82123523177698\times 10^{10}$ \wn\\
   $a_{31}$ & $ 0.18626451758590\times 10^{11}$ \wn\\
\hline
   \multicolumn{2}{|c|}{$R_\mathrm{out} < R$}\\
\hline
  ${U_\infty}$ & 0.0 \wn    \\
 ${C_6}$ &    0.1892046304$\times 10^{8}$ \wn\AA$^6$      \\
 ${C_{8}}$ &  0.5700273275$\times 10^{9}$ \wn\AA$^8$   \\
 ${C_{10}}$ & 0.1866135374$\times 10^{11}$ \wn\AA$^{10}$   \\
 ${A_{\rm ex}}$ & 0.97014411$\times 10^{4}$ \wn\AA$^{-\gamma}$   \\
 ${\gamma}$ & 5.19500    \\
 ${\beta}$ & 2.13539 \AA$^{-1}$   \\
\hline
\end{tabular*}
\end{table}

\begin{table}
\caption{Parameters of the analytic representation of the \astate\ state potential without any Born-Oppenheimer correction. The energy reference is the dissociation asymptote. Parameters with $^\ast$ are set for continuous extrapolation of the potential.  }
\label{tab:pota1}
\begin{tabular*}{0.8\columnwidth}{@{\extracolsep{\fill}}|lr|}
\hline
   \multicolumn{2}{|c|}{$R < R_\mathrm{inn}=$ 4.750 \AA}    \\
\hline
   $A^\ast$ & $-0.672898984\times 10^{3}$ \wn \\
   $B^\ast$ & $0.7735201466\times 10^{7}$  \wn \AA $^{6}$ \\
   $N_s$    & $6$\\
\hline
   \multicolumn{2}{|c|}{$R_\mathrm{inn} \leq R \leq R_\mathrm{out}=$ 12.000 \AA}    \\
\hline
    $b$ &   $-0.300$              \\
    $R_\mathrm{m}$ & 5.73392370 \AA  \\
    $a_{0}^\ast$ & $-255.016075$ \wn\\
    $a_{1}$ & $-0.83437034991917$ \wn\\
    $a_{2}$ & $ 0.20960239701879\times 10^{4}$ \wn\\
    $a_{3}$ & $-0.17090691582228\times 10^{4}$ \wn\\
    $a_{4}$ & $-0.17873986188680\times 10^{4}$ \wn\\
    $a_{5}$ & $ 0.29450770829461\times 10^{4}$ \wn\\
    $a_{6}$ & $-0.20200111692363\times 10^{5}$ \wn\\
    $a_{7}$ & $-0.35699427038012\times 10^{5}$ \wn\\
    $a_{8}$ & $ 0.59869069169566\times 10^{6}$ \wn\\
    $a_{9}$ & $-0.71054314902491\times 10^{6}$ \wn\\
   $a_{10}$ & $-0.61711835715388\times 10^{7}$ \wn\\
   $a_{11}$ & $ 0.19365507918230\times 10^{8}$ \wn\\
   $a_{12}$ & $ 0.67930591036665\times 10^{7}$ \wn\\
   $a_{13}$ & $-0.12020061749490\times 10^{9}$ \wn\\
   $a_{14}$ & $ 0.21603960091887\times 10^{9}$ \wn\\
   $a_{15}$ & $-0.63531970658436\times 10^{8}$ \wn\\
   $a_{16}$ & $-0.52391212911571\times 10^{9}$ \wn\\
   $a_{17}$ & $ 0.15913304556368\times 10^{10}$ \wn\\
   $a_{18}$ & $-0.24792546801660\times 10^{10}$ \wn\\
   $a_{19}$ & $ 0.20326032002627\times 10^{10}$ \wn\\
   $a_{20}$ & $-0.68044505933607\times 10^{9}$ \wn\\
\hline
   \multicolumn{2}{|c|}{$R_\mathrm{out} < R$}\\
\hline
  ${U_\infty}$ & 0.0 \wn    \\
 ${C_6}$ &    0.1892046304$\times 10^{8}$ \wn\AA$^6$      \\
 ${C_{8}}$ &  0.5700273275$\times 10^{9}$ \wn\AA$^8$   \\
 ${C_{10}}$ & 0.1866135374$\times 10^{11}$ \wn\AA$^{10}$   \\
 ${A_{\rm ex}}$ &$-0.97014411\times 10^{4}$ \wn\AA$^{-\gamma}$   \\
 ${\gamma}$ & 5.19500    \\
 ${\beta}$ & 2.13539 \AA$^{-1}$   \\
\hline
\end{tabular*}
\end{table}

We followed three different fit cases to get a reliable theoretical description of all observations and by comparing the results we might verify the validity of the BOA and get a limit of its precision or derive a quantitative correction to the BOA. First, we applied all data to fit potentials without any Born-Oppenheimer correction. Second, we split the data set: The spectroscopic data and the Feshbach resonances for $^{39}$K$_2$ only will determine the potentials for states \Xstate\ and \astate\ which are then used to simulate the few Feshbach resonances for $^{40}$K$_2$. Our attention was concentrated on any systematic shift seen between the observed and simulated resonances for $^{40}$K$_2$.  Third, we included correction functions as shown in Eq.~(\ref{eq:BOC2}). We again include the $^{40}$K$_2$ data and check if any improvement on the fit compared to the first case can be obtained.

For the \emph{first case} the fit of the whole data set starts with the potential approach reported in \cite{pas08}, the number of potential parameters for the intermediate regime, Eq.~(\ref{eq:uanal}), and the values of $R_{\rm inn}$ and $R_{\rm out}$ for splitting the $R$ axis into the different regions are taken from that reference. The new measurements for $v'' \geq 80$ have outer turning points in the region beyond $R_{\rm out}$ thus give additional information on the long range form. This is especially important because several different rotational states are observed. This gives a more detailed test of the potential function than the vibrational spacing of levels with the same rotational angular momentum alone. For the potential fit, the calculated hyperfine structure of these observed levels is subtracted, which was derived in the preceding iteration step with coupled channels calculation. Typically, three iteration loops, alternating potential fit and coupled channels calculation, are needed to get convergence. 

Tables \ref{tab:potx1} and \ref{tab:pota1} show the fit results of the first case without any attempt of Born-Oppenheimer correction for the two states \Xstate\ and \astate. The total fit of the Born-Oppenheimer potentials contains 8775 data points, in which Feshbach resonances are included by converting them to extrapolated rovibrational levels of those quantum states which produce the observed resonances. These are vibrational levels with $v''_{\rm X} = 84$ and $85$ for $^{39}$K$_2$ and $^{40}$K$_2$, respectively and $v''_{\rm a} = 26$ for both isotopomers. Because for $^{40}$K$_2$ a p-wave resonance was also reported, a $J''=1$ and $N''=1$ level for those vibrational states of \Xstate\ and \astate, respectively, was constructed. A normalized standard deviation of $\sigma=0.78$ was obtained for the potential fit in the iteration loop. 

In the last iteration step the fit of the 12 Feshbach resonances yields a standard deviation of $\sigma=0.76$, which is very satisfactory from the statistical point of view, but the detailed inspection of the distribution of deviations between the two isotopomers shows that for the four resonances of $^{40}$K$_2$   a  smaller scatter of the deviations is obtained than in the case of the resonances of $^{39}$K$_2$.  This asymmetry indicates that the quality of the fit might be not as good as the single number of $\sigma$ would tell. Hence, we have good reason to try the second case of our fit strategy. 

But before describing these results we would like to mention, that the photoassociation measurements on $^{39}$K$_2$ from Ref.~\cite{wan00} were incorporated in the fit and showed good consistency with the other data. On average, these measurements are at least a factor of 10 less accurate than the Feshbach data, if one transforms by the Zeeman effect the magnetic field accuracy to an uncertainty of level energies. Thus these measurements have a small weight in the evaluation. For all later calculations below, the photoassociation results are always in good agreement, thus we do not need to mention it later again.

\begin{table}
\caption{Parameters of the analytic representation of the \Xstate\ state potential with adiabatic Born-Oppenheimer correction and reference isotopomer $^{39}$K$_2$. The energy reference is the dissociation asymptote. Parameters with $^\ast$ are set for continuous extrapolation of the potential. }
\label{tab:potx2}
\begin{tabular*}{0.8\columnwidth}{@{\extracolsep{\fill}}|lr|}
\hline
   \multicolumn{2}{|c|}{$R < R_\mathrm{inn}=$ 2.870 \AA}    \\
\hline
   $A^\ast$ & $-0.263145571\times 10^{4}$ \wn \\
   $B^\ast$ & $ 0.813723194\times 10^{9}$  \wn \AA $^{12}$ \\
   $N_s$    & $12$\\
\hline
   \multicolumn{2}{|c|}{$R_\mathrm{inn} \leq R \leq R_\mathrm{out}=$ 12.000 \AA} \\
\hline
    $b$ &   $-0.40$ \\
    $R_\mathrm{m}$ & 3.92436437 \AA  \\
    $a_{0}^\ast$ &  $-4450.899484$ \wn\\
    $a_{1}$ & $ 0.30601009538111\times 10^{-1} $ \wn\\
    $a_{2}$ & $ 0.13671217000518\times 10^{5}$ \wn\\
    $a_{3}$ & $ 0.10750910095361\times 10^{5}$ \wn\\
    $a_{4}$ & $-0.20933401680991\times 10^{4}$ \wn\\
    $a_{5}$ & $-0.19385874804675\times 10^{5}$ \wn\\
    $a_{6}$ & $-0.49208915890513\times 10^{5}$ \wn\\
    $a_{7}$ & $ 0.11026639220148\times 10^{6}$ \wn\\
    $a_{8}$ & $ 0.72867339500920\times 10^{6}$ \wn\\
    $a_{9}$ & $-0.29310679369135\times 10^{7}$ \wn\\
   $a_{10}$ & $-0.12407070106619\times 10^{8}$ \wn\\
   $a_{11}$ & $ 0.40333947198094\times 10^{8}$ \wn\\
   $a_{12}$ & $ 0.13229848871390\times 10^{9}$ \wn\\
   $a_{13}$ & $-0.37617673798775\times 10^{9}$ \wn\\
   $a_{14}$ & $-0.95250413275787\times 10^{9}$ \wn\\
   $a_{15}$ & $ 0.24655585744641\times 10^{10}$ \wn\\
   $a_{16}$ & $ 0.47848257695164\times 10^{10}$ \wn\\
   $a_{17}$ & $-0.11582132109947\times 10^{11}$ \wn\\
   $a_{18}$ & $-0.17022518297651\times 10^{11}$ \wn\\
   $a_{19}$ & $ 0.39469335034593\times 10^{11}$ \wn\\
   $a_{20}$ & $ 0.43141949844339\times 10^{11}$ \wn\\
   $a_{21}$ & $-0.97616955325128\times 10^{11}$ \wn\\
   $a_{22}$ & $-0.77417530685917\times 10^{11}$ \wn\\
   $a_{23}$ & $ 0.17314133615879\times 10^{12}$ \wn\\
   $a_{24}$ & $ 0.96118849114926\times 10^{11}$ \wn\\
   $a_{25}$ & $-0.21425463041449\times 10^{12}$ \wn\\
   $a_{26}$ & $-0.78513081754125\times 10^{11}$ \wn\\
   $a_{27}$ & $ 0.17539493131251\times 10^{12}$ \wn\\
   $a_{28}$ & $ 0.37939637008662\times 10^{11}$ \wn\\
   $a_{29}$ & $-0.85271868691526\times 10^{11}$ \wn\\
   $a_{30}$ & $-0.82123523240949\times 10^{10}$ \wn\\
   $a_{31}$ & $ 0.18626451751424\times 10^{11}$ \wn\\
\hline
    $v_{0}$ & $ 0.13148609 $\wn\\
    $v_{1}$ & $ 2.08523853$ \wn\\
\hline
   \multicolumn{2}{|c|}{$R_\mathrm{out} < R$}\\
\hline
  ${U_\infty}$ & 0.0 \wn    \\
 ${C_6}$ &    0.1892652670$\times 10^{8}$ \wn\AA$^6$      \\
 ${C_{8}}$ &  0.5706799527$\times 10^{9}$ \wn\AA$^8$   \\
 ${C_{10}}$ & 0.1853042722$\times 10^{11}$ \wn\AA$^{10}$   \\
 ${A_{\rm ex}}$ & 0.90092159$\times 10^{4}$ \wn\AA$^{-\gamma}$   \\
 ${\gamma}$ & 5.19500    \\
 ${\beta}$ & 2.13539 \AA$^{-1}$   \\
\hline
\end{tabular*}
\end{table}

\begin{table}
\caption{Parameters of the analytic representation of the \astate\ state potential with adiabatic Born-Oppenheimer correction and reference isotopomer $^{39}$K$_2$. The energy reference is the dissociation asymptote. Parameters with $^\ast$ are set for continuous extrapolation of the potential.  }
\label{tab:pota2}
\begin{tabular*}{0.8\columnwidth}{@{\extracolsep{\fill}}|lr|}
\hline
   \multicolumn{2}{|c|}{$R < R_\mathrm{inn}=$ 4.750 \AA}    \\
\hline
   $A^\ast$ & $-0.6948000684\times 10^{3}$ \wn \\
   $B^\ast$ & $ 0.7986755824\times 10^{7}$  \wn \AA $^{6}$ \\
   $N_s$    & $6$\\
\hline
   \multicolumn{2}{|c|}{$R_\mathrm{inn} \leq R \leq R_\mathrm{out}=$ 12.000 \AA}    \\
\hline
    $b$ &   $-0.300$              \\
    $R_\mathrm{m}$ & 5.73392370 \AA  \\
    $a_{0}^\ast$ &  $-255.015289$ \wn\\
    $a_{1}$ & $-0.84057856111142 $ \wn\\
    $a_{2}$ & $ 0.20960112217307\times 10^{4}$ \wn\\
    $a_{3}$ & $-0.17090298954603\times 10^{4}$ \wn\\
    $a_{4}$ & $-0.17873773359495\times 10^{4}$ \wn\\
    $a_{5}$ & $ 0.29451253739583\times 10^{4}$ \wn\\
    $a_{6}$ & $-0.20200089247397\times 10^{5}$ \wn\\
    $a_{7}$ & $-0.35699524005434\times 10^{5}$ \wn\\
    $a_{8}$ & $ 0.59869055371895\times 10^{6}$ \wn\\
    $a_{9}$ & $-0.71054353363636\times 10^{6}$ \wn\\
   $a_{10}$ & $-0.61711841390175\times 10^{7}$ \wn\\
   $a_{11}$ & $ 0.19365507566961\times 10^{8}$ \wn\\
   $a_{12}$ & $ 0.67930587059121\times 10^{7}$ \wn\\
   $a_{13}$ & $-0.12020061704172\times 10^{9}$ \wn\\
   $a_{14}$ & $ 0.21603959986951\times 10^{9}$ \wn\\
   $a_{15}$ & $-0.63531969223760\times 10^{8}$ \wn\\
   $a_{16}$ & $-0.52391212820709\times 10^{9}$ \wn\\
   $a_{17}$ & $ 0.15913304648629\times 10^{10}$ \wn\\
   $a_{18}$ & $-0.24792546567713\times 10^{10}$ \wn\\
   $a_{19}$ & $ 0.20326031881106\times 10^{10}$ \wn\\
   $a_{20}$ & $-0.68044508325774\times 10^{9}$ \wn\\
\hline
    $v_{0}$ &  0.23803737 \wn\\
\hline
   \multicolumn{2}{|c|}{$R_\mathrm{out} < R$}\\
\hline
  ${U_\infty}$ & 0.0 \wn    \\
 ${C_6}$ &    0.1892652670$\times 10^{8}$ \wn\AA$^6$      \\
 ${C_{8}}$ &  0.5706799527$\times 10^{9}$ \wn\AA$^8$   \\
 ${C_{10}}$ & 0.1853042722$\times 10^{11}$ \wn\AA$^{10}$   \\
 ${A_{\rm ex}}$ & -0.90092159$\times 10^{4}$ \wn\AA$^{-\gamma}$   \\
 ${\gamma}$ & 5.19500    \\
 ${\beta}$ & 2.13539 \AA$^{-1}$   \\
\hline
\end{tabular*}
\end{table}

In the \emph{second case} we fitted only the isotopomer $^{39}$K$_2$ and used the result for calculating the four Feshbach resonances of $^{40}$K$_2$. This case corresponds to the scenario of many cold collision studies, where Feshbach resonances are observed for one isotope combination and mass scaling is applied for predicting resonances of other isotope combinations. From the statistical point of view the overall fit quality in this second case was as good as in the former one. The resonance fit of $^{39}$K$_2$ is of particular interest which yields a $\sigma =0.55$. This small value obtained by restricting the data set to a single isotopomer directly reflects the above mentioned asymmetry in the distribution of the two isotopomers. With these potential functions for \Xstate\ and \astate\ the simulation of the resonances for $^{40}$K$_2$ results in deviations of more than 0.1~mT and all are negative. Because the experimental uncertainties of these resonances are mostly below 0.01~mT this second fit cannot be accepted for the description of isotopomer $^{40}$K$_2$ from results of isotopomer $^{39}$K$_2$.

This gives us good reason to turn to our \emph{third case}, in which we introduce Born-Oppenheimer correction functions. The very sensitive point when introducing new functions is the decision how many terms, represented by the number of additional parameters, should be tried. Indeed, we are looking for small effects buried in a large set of data, which is already described fairly well in the first case, and we have to expect correlation between the potential functions and new correction functions. Thus we concentrated only on the adiabatic correction, $U_{\rm ad}$ of Eq.~(\ref{eq:BOC2}), with two expansion coefficients for the state \Xstate\ and a single one for the other, because here the isotope variation is only obtained from the resonances and thus for the asymptotic levels. We selected $^{39}$K$_2$ as the reference isotopomer, because this species had the largest data set.

The potential fit yields the same quality as for the former ones, namely a standard deviation $\sigma=0.78$, but the fit of the Feshbach resonance of both isotopomers results in an improvement with a standard deviation of $\sigma=0.59$ compared to the value 0.76 obtained in the first fit. No asymmetry is found in the present distribution of the deviations. Tables \ref{tab:potx2} and \ref{tab:pota2} show the new fit results. The correction function for \Xstate\ is a linear function in the expansion variable $\xi$, compare Eq.~(\ref{eq:xv}). We also tried a fit by using a quadratic function with two parameters, i.e. with $v_0 \times\xi^0$ and $v_2 \times\xi^2$, which will change the influence mainly for the low levels within the singlet potential. But here the potential fit was not as good as before, which indicates that the few spectral lines to deeply bound levels of the singlet ground state for the isotopic species $^{39}$K$^{41}$K also play a non-negligible role in the final evaluation.

\begin{figure}[t]
\centerline{\includegraphics[width=9cm]{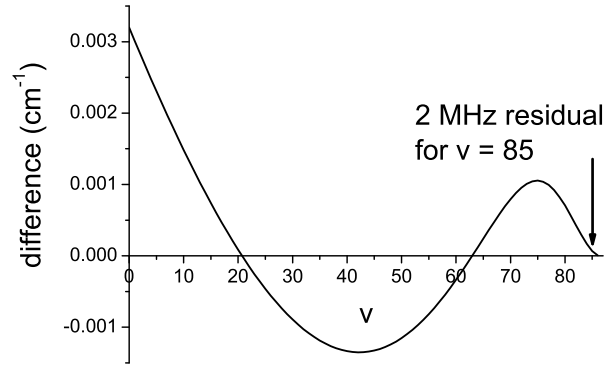}}
		 		  \caption{Difference of vibrational ladders for $^{40}$K$_2$ calculated with the potential from table \ref{tab:potx2} with adiabatic correction or setting it to zero.}
\label{fig:BOC}
\end{figure}

The influence of the derived quantitative Born-Oppenheimer correction can be seen by comparing the term energies of the obtained model to the term energies obtained from the same model with the corrections artificially set to zero. Figure~\ref{fig:BOC} shows these energy differences of the vibrational ladders for $^{40}$K$_2$. A graph for $^{39}$K$^{41}$K would almost coincide with that shown because the reduced masses of these two isotopomers are very close in value. The differences of the vibrational ladders are proportional to the expectation values of $U_{\rm ad}$ for the vibrational states $v$ of the state \Xstate. For low vibrational levels the differences are on the order of the experimental accuracy. For the asymptotic vibrational level they almost vanish as desired by fixing the asymptotic behavior according to the long range function in lowest order in Eq.~(\ref{eq:BOC2}). The actual functional form seen in Fig. \ref{fig:BOC} is far from being fixed by this study, only the magnitudes should be considered. For $v''=85$, the level which correlates to observed Feshbach resonances in $^{40}$K$_2$, the energy difference is about $2$~MHz and thus at least a factor 3 larger than the uncertainty of the resonance determination would tolerate and it corresponds to the resonance shifts for $^{40}$K$_2$ between the observed values and those calculated with the potentials derived form  for $^{39}$K$_2$ data as obtained in the second fitting case. A similar magnitude results for the levels of state \astate. In total, we can state that we reached the limit of accuracy of the Born-Oppenheimer approximation in modeling the observations on K$_2$. Small improvements of the experiments will allow to derive clearly the magnitude of the corrections to the Born-Oppenheimer approximation.

Finally, we also checked if retardation effects could alter such statement about accuracy and possible correction to the Born-Oppenheimer approximation. As described by Marinescu et al. \cite{mar94} we introduced the correction factors calculated by them for all $C_i$ coefficients in Eq.~(\ref{eq:lrexp}) and repeated the iterative fit. We got the same quality of the fit and all changes were insignificant regarding the conclusion on the magnitude of possible Born-Oppenheimer corrections and derived scattering properties, which will be collected in the next section.
%
%
%
%
\section{\label{sec:con} Conclusion}
The new measurements with a molecular beam setup close the level gap between the data from Fourier transform spectroscopy and Feshbach resonances. The simulations of the new levels with the earlier derived potentials \cite{pas08} showed that this interpolation results in predictions, which deviate by about two times the error limit from the Fourier transform spectroscopy. Precise measurements by Feshbach spectroscopy do not help significantly for obtaining better predictions in the range below those vibrational levels, which are determined by the resonances. This is important information for other studies in systems, where a combined analysis of data sets from different, not linked energy regions is performed. A direct comparison between the earlier potential curves and the present ones gives differences on the order of few times the uncertainty of the Fourier transform data, more specificly about 0.02~\wn\ and for the steep repulsive branch at short internulcear separation more, distributed over the full potential curve. Solving the Schr\"odinger equation means integration over the full range $R$ of the potential function and results for the eigenvalues only in changes of about 0.003~\wn\ or less as seen between the earlier simulated asymptotic level energies and those measured now. One should also keep in mind that the $\Lambda$-scheme of excitation applied for these new measurements gives much improved accuracy at an isolated deeply bound energy regime, i.e. around $v=41$. This fact certainly influences also the overall change in the potential function. We conclude, that having data for the full range of potential energy with no gaps will give a secure basis for reliable predictions, whereas extrapolations and even interpolations should be labeled by a remark of caution for using those. Our analysis also shows how large changes in potential energies can appear compared to the overall accuracy of level energies, namely in our case about a factor of ten.

Such limitations also restrict the chances to discuss higher order corrections in the theoretical models like our desire for searching for Born-Oppenheimer corrections. We think that we are here in a favorable situation because at least for one isotopomer, namely $^{39}$K$_2$, we have spectral information over the full range of the potential energy, including the atomic asymptote by the Feshbach spectroscopy. Comparing the different fit results from the three cases, (i) evaluation with no correction, (ii) separate evaluation of the isotopomers and finally (iii) a simultaneous evaluation including Born-Oppenheimer corrections, we found the first indications of Born-Oppenheimer corrections for the bound levels correlating to the Feshbach resonances. 

\begin{table}
\fontsize{7pt}{12pt}\selectfont
\caption{Scattering lengths (unit $a_0=0.5292$~\AA ) for the both approaches of the potentials (A: no B.-O. correction, B: with B.-O. correction) for different isotopomers of potassium. Results from experimental analysis by other authors are given: a \cite{der07}, b \cite{mod01}. }
\label{length}
\begin{tabular}{r|rrr|rrr} \hline
isotope & \multicolumn{3}{c|}{a$_{\rm singlet}$} & \multicolumn{3}{c}{a$_{\rm triplet}$}  \\
 &  A &  B& others  &  A&  B & others \\
\hline
    $39/39$ & $ 138.80$ & $138.49(12)$ & $138.90(15)^{\rm a}$ & $-33.41$ & $-33.48(18)$ & $-33.3(3)^{\rm a}$  \\
    $39/40$ & $  -2.69$ & $ -2.84(10)$ & $  $           & $-2044 $ & $ -1985(69)$ & $  $  \\
    $39/41$ & $ 113.09$ & $113.07(12)$ & $  $           & $176.57$ & $177.10(27)$ & $  $   \\
    $40/40$ & $ 104.42$ & $104.41(9) $ & $104.56(10)^{\rm a}$ & $169.18$ & $169.67(24)$ & $169.7(4)^{\rm a}$  \\
    $40/41$ & $ -54.37$ & $-54.28(21)$ & $  $           & $ 97.14$ & $ 97.39(9) $ & $  $      \\
    $41/41$ & $  85.41$ & $ 85.53(6) $ & $  $           & $ 60.27$ & $ 60.54(6) $ & $78(20)^{\rm b}$    \\
\hline
\end{tabular}
\end{table}

For characterizing the cold collision properties at zero kinetic energy we calculated the scattering lengths of the singlet and the triplet state from the results of the first and third case of fitting for the isotopic species of diatomic potassium of natural abundance. Table \ref{length} shows the results and includes also reports from other authors. The error limits are inferred by conventional methods including the covariance matrix from the nonlinear least squares fit in the linear limit; they should be interpreted as 1$\sigma$ standard deviation from the statistical distribution of the deviations within the fit. For the comparison of the different approaches we discuss only the results for $^{39}$K$_2$ and $^{40}$K$_2$, because in these cases we have direct experimental observations. All others are derived quantities only, and the respective model was used for their calculations. For the singlet state the scattering length of $^{39}$K$_2$ is different slightly outside the statistical uncertainty limit between the approach with and without Born-Oppenheimer correction. For the triplet state the same appears for the isotopomer $^{40}$K$_2$. Because one sees this difference only slightly outside the statistical uncertainty limit we cannot take this information as a conclusive observation of a correction to the Born-Oppenheimer approximation for the modeling of cold collisions, but we believe that it gives the limit of mass scaling and that any improvement in precision would need to include corrections in the theoretical description like in the form of Eq. (\ref{eq:BOC}). Cold collisions outside resonance cases are probably less influenced by possible corrections to the Born-Oppenheimer approximation and therefore mass scaling of scattering lengths might be widely applicable, but resonances themselves seem to be more sensitive, because we observed that mass scaling from $^{39}$K$_2$ to $^{40}$K$_2$ shifts resonances in the order of 0.1~mT, hence outside the most recent experimental error limits. More high precision experiments are desirable to obtain a definite conclusion. 

The results on scattering lengths are mostly in good agreement with our earlier ones \cite{pas08}, small deviations to those reported by D'Errico et al \cite{der07} could stem from the different fit qualities as already noted in \cite{pas08}. 

New precise studies of different isotopomers in other molecules are underway. We learned from Simoni \cite{simPC} about new measurements on $^{39}$K$^{87}$Rb, which they combined with earlier results on $^{40}$K$^{87}$Rb \cite{osp06,fer06b,kle07}. They concluded, that the resonances can be described applying mass-independent potentials for both states X~$^1\Sigma^+$ and a~$^3\Sigma^+$. But the derived potentials concentrate only on the asymptotic behavior to get a predicting model for ultracold collisions and the capability to precisely reproduce more deeply bound levels is lost. In this way one obtains two solutions with different ranges of validity, namely the deeply bound range \cite{pas07} or the asymptotic range. We believe that it would be valuable to repeat the procedure like in the present study of K$_2$ with full potentials and to keep the whole data set from spectroscopy and Feshbach measurements simultaneously in the fit. Otherwise, small corrections can be overseen in the artificial separation of the two solutions.

Similarly, new studies for LiK evolve. Feshbach spectroscopy was reported for the collision $^6$Li and $^{40}$K by the Innsbruck group \cite{wil08} and Fourier transform spectroscopy was continued in a Ph.D. thesis by Houssam Salami \cite{sal07} giving a large amount of data for $^6$Li$^{39}$K and $^7$Li$^{39}$K for which Salami derived corrections of the Born-Oppenheimer approximation for the deeply bound region of X~$^1\Sigma^+$. Here we have the favorable case that the reduced mass of the system is mainly determined by the light atom Li and thus the mass effect from $^6$Li to $^7$Li is larger than for heavier systems. First steps done by us with our approach already indicate very promising results, which could show the variation of the Born-Oppenheimer correction from small internuclear separations to the asymptotic range, something that could only be guessed from Fig.~\ref{fig:BOC} in the case of K$_2$.

\begin{acknowledgments}
This project was accomplished within the Sonderforschungsbereich SFB 407 supported by the Deutsche Forschungsgemeinschaft (DFG). We thank Nathan Gilfoy for carefully reading the manuscript.
\end{acknowledgments}

\end{document}